\documentclass[showpacs,aps,nofootinbib,twocolumn,floatfix]{revtex4}

\usepackage{graphicx}
\usepackage{latexsym}

\sloppypar

\newcommand{\refe}[1]{(\ref{#1})}

\newcommand{\be}{\begin{displaymath}}
\newcommand{\ee}{\end{displaymath}}
\newcommand{\bel}{\begin{equation}}
\newcommand{\eel}{\end{equation}}
\newcommand{\bea}{\begin{eqnarray}}
\newcommand{\eea}{\end{eqnarray}}
\newcommand{\ba}[1]{\begin{array}{#1}}
\newcommand{\ea}{\end{array}}
\newcommand{\ra}{\rightarrow}

\newcommand{\mrm}{\mathrm}
\newcommand{\mcm}{{\mathcal M}}

\newcommand{\mch}{{\mathcal H}}
\newcommand{\intl}{\int\limits}

\newcommand{\JPRC}[3]{Phys. Rev. C {\bf #1}, #2 (#3)}
\newcommand{\JPRD}[3]{Phys. Rev. D {\bf #1}, #2 (#3)}
\newcommand{\JNP}[3]{Nucl. Phys. {\bf #1}, #2 (#3)}

\newcommand{\JEPJA}[3]{Eur. Phys. J. A {\bf #1}, #2 (#3)}

\begin{document}

\title{What is the Correct $\pi^- p \ra \omega n$ Cross Section at
  Threshold?\footnote[2]{Work supported by DFG and GSI Darmstadt.}}  

\author{G. Penner}
\email{gregor.penner@theo.physik.uni-giessen.de}
\author{U. Mosel} 
\affiliation{Institut f\"ur Theoretische Physik, Universit\"at Giessen, D-35392
Giessen}

\begin{abstract}
The $\pi^- p \ra \omega n$ threshold cross section of
refs. \cite{binnie,keyne,karami} resisted up to now a 
consistent theoretical description in line with experiment, 
mainly caused by too large Born contributions. This led to a 
discussion in the literature about the correctness of the extraction
of these data points. We show that the extraction method used in these
references is indeed correct and that there is no reason to doubt
the correctness of these data. 
\end{abstract}


\maketitle

\section{Introduction}

In the 70's a series of experiments \cite{binnie,keyne,karami} was
performed to measure the $\pi^- p \ra \omega n$ cross section just above 
threshold. These data resisted up to now a consistent theoretical
description, mainly caused by too large Born 
contributions \cite{klingl}. As a consequence these diagrams were either
neglected \cite{post,friman} or suppressed by very soft formfactors
\cite{titov}. These findings motivated a discussion in the literature
about the experimentalists' way to extract the two-body cross section 
\cite{hanhart99} and readjustments of the published 
$\pi^- p \ra \omega n$ cross section data were
performed \cite{sibi,titov,hanhart01}. 

The cause of the discussion is the experimentalists' unusual method 
to cover the full range of the $\omega$ spectral function. An
integration over at least one kinematical variable is necessary to
make sure that all pion triples with invariant masses around
$m_\omega$ are taken into account, so that the $\omega$ spectral
function with a width of $8$ Mev is well covered. Instead of fixing
the incoming pion momentum and integrating out the invariant mass of
the pion triples directly, the authors of \cite{binnie,keyne,karami}
fixed the outgoing neutron laboratory momentum and angle and performed
an integration over the incoming pion momentum. 

Led by the observation that the cross sections of
\cite{binnie,keyne,karami} result in a
hard-to-understand energy dependence of the transition matrix element
the authors of \cite{hanhart99} claimed that due to the
experimental method just described the count rates covered only a
fraction of the $\omega$ spectral function. As a consequence, the
authors of \cite{hanhart99,sibi,titov} advocated that the two-body total
cross sections given in \cite{karami,landolt} should be
modified. Imposing this modification of the threshold cross section,
in \cite{hanhart99} a practically constant transition matrix element 
up to $1.74$ GeV corresponding to an $S_{11}$ or $D_{13}$
\cite{hanhartprivate} (since the pion momentum is almost constant in
this region) wave production mechanism (in the usual $\pi N$ notation,
i.e. an $IJ^P=\frac{1}{2}\frac{1}{2}^-$ or
$IJ^P=\frac{1}{2}\frac{3}{2}^-$, resp., partial wave) was deduced. 

\begin{figure}[t]
  \begin{center}
    \parbox{75mm}{\includegraphics[width=75mm]{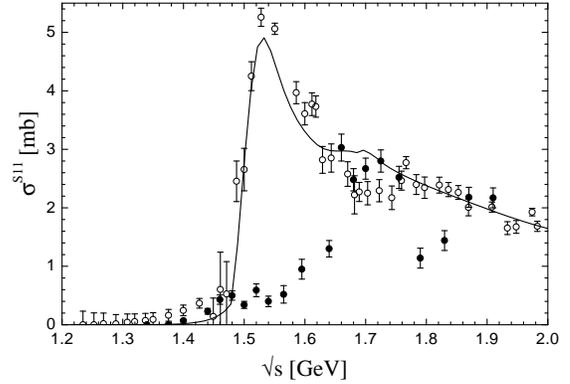}}
    \caption{$S_{11}$ inelastic partial wave cross section of
      $\pi N \ra \pi N$ as deduced from SM00 \cite{SM00}
      $\circ$ and $\pi N \ra 2 \pi N$ partial wave cross section as
      deduced by \cite{manley} $\bullet$. The curve gives the
      result of a coupled-channel analysis \cite{moretocome}.
      \label{s11inel} }  
  \end{center}
\end{figure}

This modification, however, immediately raises another
problem. Taking into account this ``spectral function correction''
increases the $\omega N$ cross section for $1.72 \leq \sqrt s \leq
1.74$ up to $\sigma \geq 3$ mb, which is in contradiction to the
inelasticity deduced from $\pi N \ra \pi N$ partial wave analyses
(e.g. SM00 \cite{SM00}). 
As can be seen in fig. \ref{s11inel} 
the $\pi N \ra \pi N$
inelasticity in the $S_{11}$ partial wave is already saturated by $\pi
N \ra 2 \pi N$ in this energy region, i.e. an 
additional $\omega N$ contribution of $\sigma^{S_{11}} \geq 3$ mb
would lead to a gross overestimate of the inelasticity: 
\bea
\sigma^{S_{11}}_{in} &\geq& 
\sigma^{S_{11}}_{2\pi N} + \sigma^{S_{11}}_{\omega N} 
\nonumber \\
&\geq& 
2.5 \; \mrm{mb} + 3 \; \mrm{mb} \gg 
2.4 \; \mrm{mb} \approx \sigma^{S_{11}}_{in} (\mrm{SM00}) \; .
\eea
The same argument holds for the $D_{13}$ wave (see fig. 5 in
\cite{moretocome}). 

Because of this inconsistency of the newly extracted cross sections
with existing inelasticities and because of the importance of the $\pi
N \ra \omega N$ cross section for both unitary models analyzing
reactions on the nucleon in the c.m. energy range $1.7 \leq \sqrt s \leq
2.0$ GeV \cite{moretocome} and in-medium models of vector mesons
\cite{post,friman97,klingl99} we reanalyze the extraction method used
in refs. \cite{binnie,keyne,karami} by presenting a complete
derivation (given in parts in \cite{binnie}) of the relation between
the experimental count rates and the extracted two-body cross section for
$\pi N \ra \omega N$. 

\section{Count Rates and Two-Body Cross Sections}

The experimental count rates are given by (B3) (we label all equations
to be found in \cite{binnie} with the letter B and those in 
\cite{hanhart99} by the letter H): 
\bel
\bar N = N_H \intl_{\Delta \Omega_L} \intl_{\Delta \mrm p'_L}
\frac{\mrm d^3 \sigma}{\mrm d \Omega_L \mrm d \mrm
  p'_L} \mrm d \Omega_L \mrm d \mrm p'_L \; ,
\label{binnie3}
\eel
where the cross section $\sigma$ describes the process $\pi^- p \ra n
\pi^+ \pi^- \pi^0$. All variables are taken in the c.m. system, unless
they are denoted by the label L (laboratory frame). $p$ ($k$) and $p'$
($q$) denote the incoming proton and outgoing neutron (incoming $\pi$
and outgoing $\omega$) four 
momenta. Absolute values of three momenta are denoted by upright
letters, $s$ and $t$ are the usual Mandelstam variables. $\Omega_L$
is the neutron laboratory solid angle and $N_H$ the number of
target particles per unit area. The integral ranges in \refe{binnie3}
refer to the binning of the count rates, i.e. they are the integrals
to be performed for averaging over the experimental resolution
intervals $\Omega_L \pm \Delta \Omega_L$ and $\mrm p'_L \pm \mrm
\Delta \mrm p'_L$ and are not related to an integration over the
$\omega$ spectral function. This will become clearer below
(cf. eqn. \refe{binniecounts}). The  kinematics of the reaction are
extracted from the center values of these intervals. Using 
\bea
t = (p_L-p_L')^2 &=& m_p^2 + m_n^2 - 2 m_p {E_n}_L 
\nonumber \\
&=& 
m_p^2 + m_n^2 - 2 m_p \sqrt{m_n^2 + {\mrm p'_L}^2}
\label{mandeltlab}
\eea
and
\bea
(q_L-p_L)^2 &=& (k_L-p'_L)^2 \; , \; \;
\; {E_\omega}_L = {E_\pi}_L + m_p - {E_n}_L 
\nonumber \\
\Longrightarrow \;
q^2 &=& m_\pi^2 + m_n^2 + m_p^2 + 
\label{q2} \\
&& 2 (m_p {E_\pi}_L - m_p {E_n}_L -
{E_\pi}_L  {E_n}_L + \mrm k_L \mrm p'_L x_L )
\nonumber
\eea
($x_L =\cos \vartheta_L$) one finds
\be
\bar N = \frac{N_H}{2\pi} \intl_{\Delta \Omega_L} \intl_{\Delta
  \mrm p'_L} 
\frac{\mrm d^2 \sigma}{\mrm d t d \sqrt{q^2}} J_P \mrm d
\Omega_L \mrm d \mrm p'_L \; ,
\ee
assuming the cross section is independent of the neutron azimutal
angle $\varphi$. The Jacobian $J_P$ is given by
\be
J_P = \left| 
\begin{array}{cc} \mrm d \sqrt{q^2}/\mrm d \mrm p'_L & \mrm d
  \sqrt{q^2}/\mrm d x_L \\ 
\mrm d t/\mrm d \mrm p'_L & \mrm d t/\mrm d x_L \end{array}
\right| = \frac{2 m_p \mrm k_L {\mrm p'_L}^2}{\sqrt{q^2} {E_n}_L} 
\; ,
\ee
because $\mrm d t / \mrm d x_L = 0$. Since in the actual experiment
the time of flight $\tau_L$ of the neutron over a distance $d$, 
\be
\tau_L = \frac{1}{\beta_L}\frac{d}{c}
\ee
with the velocity $\beta_L = \mrm p'_L / {E_n}_L$, is measured -- not
its three-momentum --, the count rate is reexpressed in terms of the
time of flight: 
\be
\bar N = \frac{N_H}{2\pi} \intl_{\Delta \Omega_L} \intl_{\Delta
  \tau_L} 
\frac{\mrm d^2 \sigma}{\mrm d t \mrm d \sqrt{q^2}} J_\tau \mrm d
\Omega_L \mrm d \tau_L \; ,
\ee
with
\be
J_\tau = J_P \frac{\mathrm d \mathrm p'_L}{\mathrm d \tau_L} \; .
\ee
The factor linking the Jacobians is
\be
\frac{\mathrm d \mathrm p'_L}{\mathrm d \tau_L} = 
\frac{\mathrm d \mathrm p'_L}{\mathrm d \beta_L} \frac{\mathrm d
  \beta_L}{\mathrm d \tau_L} = 
\frac{{\mathrm p'_L}^2 {E_n}_L}{m_n^2} \frac{c}{d}
 \; 
\mbox{ because of }
 \;
\frac{\mrm d \beta_L}{\mrm d \mrm p'_L} = \frac{m_n^2}{{E_n}_L^3}
 \; .
\ee

We now relate this count rate to the two-body cross section
\cite{itzykson} 
\bel
\frac{\mrm d \sigma^{2b}}{\mrm d t} = 
\frac{1}{64 \pi s \mrm k^2} 
\frac{1}{2} \sum_{\lambda_p, \lambda_n, \lambda_\omega} 
|\mcm (s,t)|^2 
\label{2bodyx}
\eel
of $\pi^- p \ra \omega n$, assuming a stable $\omega$. In order to do
so, we deviate from the derivation in \cite{binnie} and start with the
general cross section formula for $\pi N \ra 3+4+5+\dots+l$ in the c.m.
system \cite{itzykson}:  
\bea
d \sigma &=& \frac{1}{4 \mrm k \sqrt s} \frac{1}{2} \sum
|\tilde \mcm|^2 (2 \pi)^4 
\prod_{j=3}^l \frac{\mrm d^3 k'_j}{(2\pi)^3 2 E'_j} 
\nonumber \\
&& \hspace{18mm} \times
\delta^4(p+k-\sum\nolimits_{j=3}^l k'_j) 
\; , 
\label{totitzy}
\eea
where the sum stands for summing over initial and final 
spins. Here (for simplicity, we assume that the $\omega$ only decays
into $3$ pions: $\pi^- p \ra n \omega \ra n \pi^+ \pi^- \pi^0$), the
matrix element reads:
\be
\tilde \mcm = \mcm_{\pi^- p \ra \omega n}^\mu D_{\mu \nu}^\omega (q^2) 
\mch_{\omega \ra 3 \pi}^\nu
\ee
with $D^\omega_{\mu \nu} (q^2) = \sum_{\lambda_\omega}
\varepsilon^\dagger_\mu (\lambda_\omega) \varepsilon_\nu (\lambda_\omega)
\Delta_\omega(q^2)$ and $\Delta_\omega(q^2) = (q^2 - m_\omega^2 + i
\sqrt{q^2} \Gamma_{\omega \ra 3 \pi})^{-1}$. Thus, in $|\tilde
\mcm|^2$ a sum over $\lambda_\omega$ and $\lambda'_\omega$
appears. However, since the $\omega$ decay amplitude
$\varepsilon_{\lambda_\omega} \cdot \mch_{\omega \ra 3 \pi}$
can be decomposed into spherical harmonics $Y_{1\lambda_\omega}$ and
the outgoing  pion angles are integrated out, there are only
contributions for $\lambda_\omega = \lambda'_\omega$. To introduce the 
$\omega$ spectral function $\rho_\omega(q^2)$ in \refe{totitzy}, we
note that we can evaluate the decay $\varepsilon_{\lambda_\omega}
\cdot \mch_{\omega \ra 3 \pi}$ in the $\omega$ rest frame, which is
hence independent of the polarization $\lambda_\omega$. Then the width 
of the $\omega$ is given by
\bea
\Gamma_{\omega \ra 3 \pi} (q^2) &=&
\frac{1}{2 \sqrt{q^2}} 
\int \prod_{j=4}^6 \frac{\mrm d^3 k'_j}{(2\pi)^3 2 E'_j} 
\left| \varepsilon_{\lambda_\omega}^\mu \cdot \mch_\mu
\right|^2 
\nonumber \\
&& \hspace{14mm} \times
(2 \pi)^4 \delta^4(q-\sum\nolimits_{j=4}^6 k'_j) ,
\eea
valid for any $\lambda_\omega$, and related to the $\omega$ spectral
function in the following way: 
\bea
\rho_\omega (q^2) &=& 
- \frac{1}{\pi} \mrm{Im} \Delta_\omega (q^2) \nonumber \\
&=& \frac{1}{\pi} \left|\Delta_\omega (q^2)\right|^2 \sqrt{q^2}
\Gamma_\omega (q^2) 
\nonumber \\
&=& \frac{1}{\pi} \left|\Delta_\omega (q^2)\right|^2 \frac{1}{2} 
\int \prod_{j=4}^6 \frac{\mrm d^3 k'_j}{(2\pi)^3 2 E'_j} 
|\varepsilon_{\lambda_\omega}^\mu \cdot \mch_\mu |^2 \nonumber \\
&& \hspace{26mm} \times
(2 \pi)^4 \delta^4(q-\sum\nolimits_{j=3}^6 k'_j)
\; .
\nonumber
\eea
Now, we can rewrite the cross section of eqn. \refe{totitzy} by
introducing $1=\int \mrm d^4 q \delta^4(q-\sum_{j=3}^6 k'_j)$ and
using the spectral function:
\begin{widetext}
\begin{eqnarray}
\mrm d^2 \sigma &=& \frac{1}{4 \mrm k \sqrt s} 
\frac{1}{2} \sum_{\lambda_p, \lambda_n, \lambda_\omega} 
\int \left|\mcm \cdot \varepsilon_{\lambda_\omega}\right|^2 
(2 \pi)^4 \delta^4(p+k-p'-q) 
\frac{\mrm d^3 p'}{(2\pi)^3 2 E_n}
\frac{2 \pi}{(2\pi)^4} \rho_\omega (q^2) \mrm d^4 q 
\nonumber \\
&=& \left. \frac{1}{4 \mrm k \sqrt s} \frac{1}{(2 \pi)^2} 
\frac{1}{2} \sum_{\lambda_p, \lambda_n, \lambda_\omega} 
\int \left|\mcm \cdot \varepsilon_{\lambda_\omega}\right|^2 
\frac{\mrm d^3 p'}{2 E_n}
\rho_\omega (q^2) \right|_{q = p+k-p'}
\nonumber \\
&=& \left. \frac{1}{8 \mrm k \sqrt s} \frac{1}{2 \pi} 
\frac{1}{2} \sum_{\lambda_p, \lambda_n, \lambda_\omega} 
\left|\mcm \cdot \varepsilon_{\lambda_\omega}\right|^2 
\frac{\mrm{p'}^2 \mrm d \mrm{p'}}{E_n} \mrm d x
\rho_\omega (q^2) \right|_{q = p+k-p'}
\label{intermediate} \\
&=& \left. \frac{\mrm p'}{32 \pi \sqrt s \mrm k^2} 
\frac{1}{2} \sum_{\lambda_p, \lambda_n, \lambda_\omega} 
\left|\mcm \cdot \varepsilon_{\lambda_\omega}\right|^2 
\frac{\mrm d \mrm p' \mrm d t}{E_n}
\rho_\omega (q^2) \right|_{q = p+k-p'}
\nonumber \\
&\stackrel{\refe{2bodyx}}=& \left. 2 \sqrt{q^2} 
\frac{\mrm d \sigma^{2b}}{\mrm d t} \rho_\omega (q^2) \mrm d t
\mrm d \sqrt{q^2} \right|_{q = p+k-p'} \; ,
\nonumber
\end{eqnarray}
\end{widetext}
where we have used
\bea
t = (p-p')^2 &=& m_p^2+m_n^2 - 2 ( E_p E_n - \mrm k \mrm p' x) 
\label{mandeltcms} \\
\Longrightarrow \;
\left| \frac{\mrm d t}{\mrm d x} \right| &=& 2 \mrm k \mrm p' \; ,
\nonumber
\eea
and 
\bea
E_n &=& \frac{s+m_n^2-q^2}{2 \sqrt s} \label{enerneutron} \\
\Longrightarrow \;
\frac{\mrm d \mrm p'}{\mrm d \sqrt{q^2}} &=& 
\frac{\mrm d \mrm p'}{\mrm d E_n} 
\frac{\mrm d E_n}{\mrm d \sqrt{q^2}} = 
\frac{\sqrt{q^2} E_n}{\mrm p' \sqrt s} \; .
\nonumber
\eea
Thus we have
\be
\frac{\mrm d^2 \sigma}{\mrm d t \mrm d \sqrt{q^2}}
= \left. 2 \sqrt{q^2} 
\frac{\mrm d \sigma^{2b}}{\mrm d t} 
\rho_\omega (q^2) \right|_{q = p+k-p'}
\nonumber
\ee
and finally for the experimental count rate
\bea
\hspace*{-4mm}
\bar N = \frac{N_H}{2\pi} \intl_{\Delta \Omega_L} \intl_{\Delta
  \tau_L} 
\left. 2 \sqrt{q^2} \frac{\mrm d \sigma^{2b}}{\mrm d t} 
\rho_\omega (q^2) \right|_{q = p+k-p'} \hspace{-8mm}
J_\tau \mrm d \Omega_L \mrm d \tau_L . &&
\label{countfixedener}
\eea
To eliminate the $\omega$ spectral function, the experimentalists
\cite{binnie,keyne,karami} performed an integration over the incoming
pion three-momentum $\mrm k_L$ by summing over all beam settings,
which can be transformed into an integration over the $\omega$
four-momentum squared:
\bea
N &=& \int \mrm d \mrm k_L \bar N \nonumber \\
&=& \frac{N_H}{2\pi} \int \mrm d \mrm k_L 
\intl_{\Delta \Omega_L} \intl_{\Delta \tau_L}
2 \sqrt{q^2} \frac{\mrm d \sigma^{2b}}{\mrm d t} 
\rho_\omega (q^2) J_\tau \mrm d \Omega_L \mrm d \tau_L 
\nonumber \\
&=& \frac{N_H}{2\pi} 
\intl_{\Delta \Omega_L} \intl_{\Delta \tau_L}
\frac{\mrm d \sigma^{2b}}{\mrm d t} 
J_k J_\tau \mrm d \Omega_L \mrm d \tau_L  \; ,
\label{binniecounts}
\eea
where we have used the normalization of the spectral function $\int
\rho_\omega(q^2) \mrm d q^2 = 1$ and the assumption that the matrix
element in eqn. \refe{2bodyx} only varies slightly around the peak of
the $\omega$ spectral function\footnote{The approximate constancy of
  the matrix element is the basic assumption for extracting the two-body
  cross section from \textbf{any} experiment dealing with decaying
  final state particles.} for fixed $t$ corresponding to fixed $\mrm
p'_L$ via eqn. \refe{mandeltlab}\footnote{Remember that the integral
  range $\Delta \tau_L$ in eqn. \refe{binniecounts} corresponds to the
  experimental resolution.}. In the first line, due to four-momentum
conservation, the $\omega$ momentum is restricted to $q=p+k-p'$. The
Jacobian $J_k$ is given by  
\be
J_k \equiv \frac{\mrm d \mrm k_L}{\mrm d \sqrt{q^2}}
\stackrel{\refe{q2}}= 
\frac{\sqrt{q^2}}{\frac{\mrm k_L}{{E_\pi}_L} \left( m_p - {E_n}_L
      \right) + \mrm p'_L x_L} \; .
\ee
In the derivation of eqn. \refe{binniecounts} two more assumptions
enter that are checked in the following: 
\begin{itemize}
\item{Sufficient coverage of the $\omega$ spectral function for all
    kinematics extracted, even at low $\mrm p'$ values: \\
    The incoming pion lab momentum range was $\mrm k_L \in
    [1.04,1.265]$ GeV, which translates into a c.m. energy range of
    $\sqrt s \in [1.6938,1.8133]$ GeV. Using $\mrm p' \in [0.03,0.21]$
    GeV, this leads to the following ranges for the upper and lower
    limits in the $q^2$ integration: $q^2_+ \in [0.822^2,0.869^2]$
    GeV$^2$ and $q^2_- \in [0.7^2,0.753^2]$ GeV$^2$. Even in the worst
    case the integration extends over at least $7$ half widths
    $\Gamma_\omega / 2$ on either side of $m_\omega^2$ and thus covers
    more than $92\%$ of the spectral function.}
\item{Constancy of the product of the Jacobians $J_k J_\tau$: \\
    By using eqns. \refe{mandeltlab}, \refe{q2}, \refe{mandeltcms},
    and \refe{enerneutron} one can easily show that the product
    \be
    J_k J_\tau = 
    \frac{2 m_p \mrm k_L {\mrm p'_L}^4}
    {m_n^2 \left[ \frac{\mrm k_L}{{E_\pi}_L} \left( m_p - {E_n}_L
        \right) + \mrm p'_L x_L \right]} 
    \frac{c}{d}
    \nonumber
    \ee
    varies for fixed $t$ by less than $2.5$ percent in the interval
    $\sqrt{q^2} \in [m_\omega - 2 \Gamma_\omega, \; m_\omega + 2
    \Gamma_\omega]$ for all kinematics considered in the experiment.}
\end{itemize}
Since the countrate of eqn. \refe{binniecounts} is identical to the 
one given between eqns. (B9) and (B10) in \cite{binnie}, the method
used in refs. \cite{binnie,keyne,karami} and correspondingly their
two-body cross sections are correct.

\begin{figure}[t]
  \begin{center}
    \parbox{75mm}{\includegraphics[width=75mm]{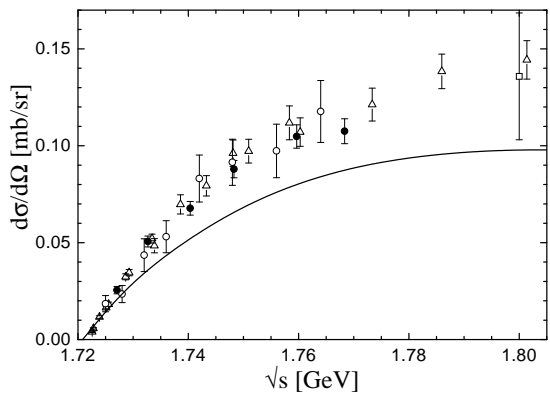}}
    \caption{Differential cross section for backward $\omega$ c.m.
      angles. The data points are from $\bullet$:
      \cite{binnie}, $\triangle$: \cite{keyne},
      $\circ$: \cite{karami}, and $\Box$:
      \cite{danburg}. The curve gives the result of a
      coupled-channel analysis
      \cite{moretocome}.\label{difbackward}}  
  \end{center}
\end{figure}

It is important to notice that in the data analysis both $\mrm p'_L$
\textbf{and} 
$x_L$ are needed to fix the kinematics of the measured
events. During the count rate corrections (flux normalization in
dependence on $\mrm k_L$, background subtraction via missing mass
spectra), for each $\mrm k_L$ beam setting the measured $\mrm p'_L$ and
$x_L$ translate into $\sqrt{q^2}$ (see \refe{q2}) and can also be
Lorentz transformed into their c.m. values $\mrm p' = \mrm
p'({E_\pi}_L, \mrm p'_L, x_L)$ and $x = x({E_\pi}_L, \mrm p'_L,
x_L)$. The events can now be regrouped in $\mrm p'$, 
$\sqrt{q^2}$, and $x$ intervals. Then, after having  
performed the integration over all $\sqrt{q^2}$ as in
\refe{binniecounts}, the translation of a given $\mrm p'$ into $\sqrt
s$ can only be done by assuming that the main contribution to the
corrected count rates comes from around the peak of the omega spectral
function $q^2 \sim m_\omega^2$.

The main assertion of ref. \cite{hanhart99}, manifested in eqn. (H4),
is that instead of \refe{binniecounts} only a fraction of the cross
section for the production of an unstable particle had been measured
in \cite{binnie,keyne,karami}. This fraction is determined by
translating the experimental $\mrm {p'}$ binning intervals given in
\cite{karami} into interval bounds for the integration over the
$\omega$ spectral function. Equation (H4), which is used for the cross
section corrections in \cite{sibi,titov,hanhart01}, is identical to
the third line of eqn. \refe{intermediate} under the assumption that
$\mrm {p'}$ is bound to the $\mrm {p'}$ binning intervals. However, as
pointed out above, in an experimental event ${E_\pi}_L$, $\mrm p'_L$,
and $x_L$ are fixed and hence eqn. \refe{countfixedener} has to be
applied to the experimental count rate for a fixed pion momentum. 

The experimental integration over the incoming pion momentum is 
introduced in ref. \cite{hanhart99} only in the subsequent discussion
between eqns. (H9) and (H10). In the paragraph following eqn. (H16)
the authors of ref. \cite{hanhart99} argue that the range of this
integral is narrowed due to the $\mrm {p'}$ binning (as in (H4)). The
$\omega$ mass is thus allowed to vary only in the interval given by
the $\mrm {p'}$ interval ranges for a fixed pion momentum. But as
shown above, fixing $\mrm {p'}$ only fixes the incoming pion
momentum if one assumes a specific $\omega$ mass. Hence the pion
momentum integration performed in the data analysis indeed translates
into an $\omega$ mass integration only bounded by the pion momentum
range and thus leads to eqn. \refe{binniecounts}. This relation
between the neutron momentum $\mrm {p'}$, the pion momentum $\mrm k$,
and the $\omega$ mass were thus treated improperly in
ref. \cite{hanhart99}. 

The second correction factor extracted in ref. \cite{hanhart99} due to
the neutron momentum binning of $\Delta \mrm p' = 10$ MeV is nothing
but the result of averaging the third line of
eqn. \refe{intermediate} over the neutron c.m. momentum: 
$(\Delta \mrm p')^{-1} \int_{\mrm p'-\Delta \mrm p'}^{\mrm p'+\Delta
  \mrm p'} {\tilde {\mrm p}}^2 \mrm d \tilde {\mrm p} = {\mrm p'}^2 +
(\Delta \mrm p')^2/12$ (cf. (H10)). This differs at most (``worst''
case: $\mrm p' = 30$ MeV) by 1 percent from ${\mrm p'}^2$ and is
therefore negligible. 

\begin{figure}[t]
  \begin{center}
    \parbox{75mm}{\includegraphics[width=75mm]{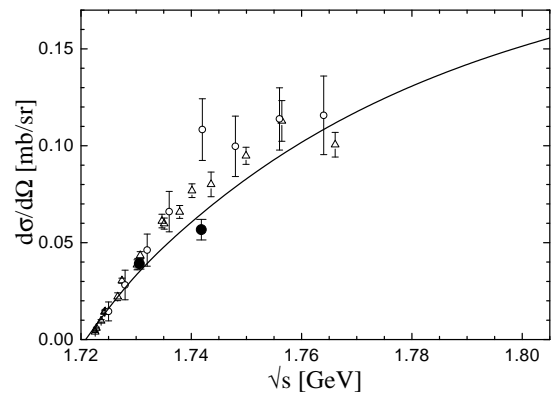}}
    \caption{Differential cross section for forward $\omega$ c.m.
      angles. For the notations see
      fig. \ref{difbackward}.\label{difforward}}
  \end{center}
\end{figure}

\begin{figure}[t]
  \begin{center}
    \parbox{75mm}{\includegraphics[width=75mm]{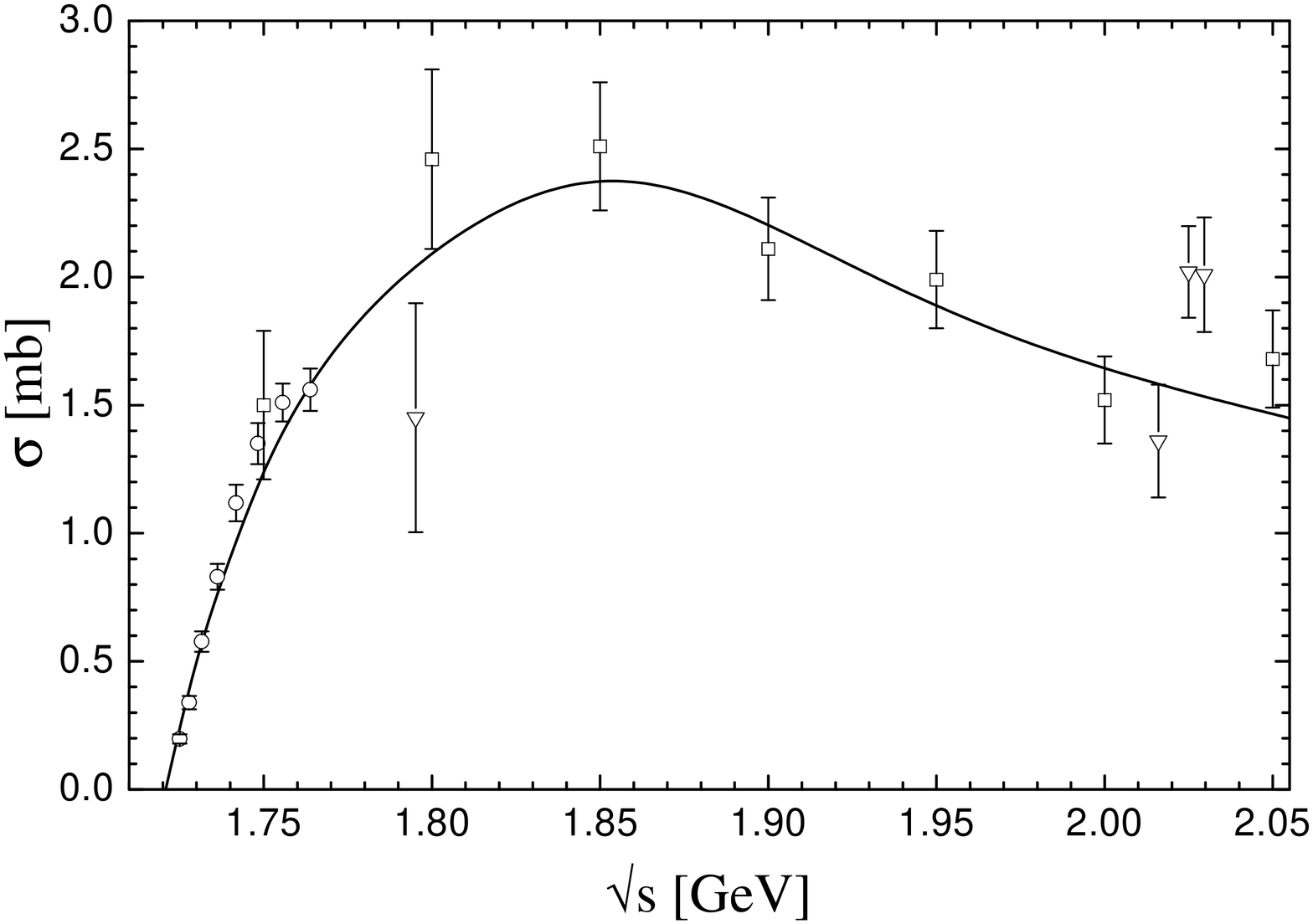}}
    \caption{$\pi N \ra \omega N$ total cross section. Data are from 
      $\circ$: \cite{karami}, $\bigtriangledown$:
      \cite{landolt}, $\Box$: \cite{danburg}. The
      curve gives the result of a coupled-channel analysis
      \cite{moretocome}.\label{totcross}}   
  \end{center}
\end{figure}

Furthermore, it is obvious from figs. \ref{difbackward} and
\ref{difforward} 
that the differential $\omega N$ data from all three references
\cite{binnie,keyne,karami}\footnote{The total cross sections given in
  refs. \cite{binnie,keyne} are actually angle-differential cross
  sections (mostly at forward and backward neutron c.m. angles)
  multiplied with $4\pi$.} are completely in line with each other and
also with ref. \cite{danburg}\footnote{The differential cross sections
  are extracted from the corrected cosine event distributions given in
  ref. \cite{danburg} with the help of their total cross sections.}. 
The same holds true for the total cross sections of ref. \cite{karami} 
in comparison with other experiments\footnote{Note, that all other
  experiments measured $\pi^+ n \ra \omega p$.}, see
fig. \ref{totcross}. 
There is, therefore, no reason to hypothesize -- as in
\cite{hanhart01} -- that the formalism developed in \cite{binnie}
could have been used incorrectly in \cite{keyne} and \cite{karami}. 

In this context we stress one more point. Very close to threshold,
the two-body cross section $\pi^- p \ra \omega n$ extracted from
experimental count rates could be influenced by the strong $\pi N$
interaction for slow pions stemming from $\omega \ra 3 \pi$. However,
this point was checked in ref. \cite{keyne} by also looking at $\omega
\ra \pi^0 \gamma$; they did not find any deviations between the two
ways of extraction.

\section{Summary and Conclusion}

We have shown that the extraction method presented in \cite{binnie}
and also used in \cite{keyne,karami} is indeed correct. 
There is no reason to doubt the correctness of the data presented in 
these references; they are in line with each other and also with other 
experimental data. The reanalysis of the $\omega N$-production data in 
ref. \cite{hanhart99}, on which the theoretical descriptions of
refs. \cite{sibi,titov} are based, as well as the speculations in
ref. \cite{hanhart01} thus lack any basis. In ref. \cite{moretocome}
we show how the $\pi^- p \ra \omega n$ cross section at threshold can
be understood in a coupled-channel analysis.

\end{document}